\documentclass[journal]{IEEEtran}

\usepackage{amsmath,amsfonts}
\usepackage{algorithm}
\usepackage{array}
\usepackage[caption=false,font=normalsize,labelfont=sf,textfont=sf]{subfig}
\usepackage{textcomp}
\usepackage{stfloats}
\usepackage{url}
\usepackage{verbatim}
\usepackage{graphicx}
\usepackage{cite}
\usepackage[framemethod=TikZ]{mdframed}
\usepackage{algpseudocode}
\usepackage{multirow}

\usepackage{pifont}
\usepackage{threeparttable}
\usepackage{booktabs}
\usepackage{ragged2e}
\usepackage{makecell}
\usepackage[justification=centering]{caption}
\usepackage{comment}

\usepackage{soul,color}

\begin{document}

\title{Dullahan: Stealthy Backdoor Attack against Without-Label-Sharing Split Learning}

% \author{~\IEEEmembership{Member, IEEE}
%\author{Anonymous Author(s)}

\author{Yuwen Pu, Zhuoyuan Ding, Jiahao Chen, Chunyi Zhou, Qingming Li, Chunqiang Hu, Shouling Ji
\thanks{Yuwen Pu, Zhuoyuan Ding, Jiahao Chen, Chunyi Zhou, Qingming Li and Shouling Ji are with the College of Computer Science and Technology at Zhejiang University, Hangzhou, Zhejiang, 310027, China. E-mail: \{yw.pu, 22251031, xaddwell, zhouchunyi, liqm, sji\}@zju.edu.cn.}
\thanks{Chunqiang Hu is with the School of Big Data \& Software Engineering, Chongqing University, Chongqing 400030, China, E-mail: chu@cqu.edu.cn.}
\thanks{Yuwen Pu and Zhuoyuan Ding are the co-first authors.}
}

% Remember, if you use this you must call 
% \IEEEpubidadjcol 
% in the second column for its text to clear the IEEEpubid mark.

\maketitle
% \IEEEtitleabstractindextext{
\begin{abstract}
As a novel privacy-preserving paradigm aimed at reducing client computational costs and achieving data utility, split learning has garnered extensive attention and proliferated widespread applications across various fields, including smart health and smart transportation, among others. While recent studies have primarily concentrated on addressing privacy leakage concerns in split learning, such as inference attacks and data reconstruction, the exploration of security issues (e.g., backdoor attacks) within the framework of split learning has been comparatively limited. Nonetheless, the security vulnerability within the context of split learning is highly posing a threat and can give rise to grave security implications, such as the illegal impersonation in the face recognition model. Therefore, in this paper, we propose a \textbf{S}tealthy \textbf{B}ackdoor \textbf{A}ttack \textbf{S}trategy (namely \texttt{Dullahan}) tailored to the without-label-sharing split learning architecture, which unveils the inherent security vulnerability of split learning. We posit the existence of a potential attacker on the server side aiming to introduce a backdoor into the training model, while exploring two scenarios: one with known client network architecture and the other with unknown architecture. Diverging from traditional backdoor attack methods that manipulate the training data and labels, we constructively conduct the backdoor attack by injecting the trigger embedding into the server network. Specifically, our \texttt{Dullahan} achieves a higher level of attack stealthiness by refraining from modifying any intermediate parameters (e.g., gradients) during training and instead executing all malicious operations post-training. Finally, we conducted extensive experiments on $3$ different models, $3$ datasets, and various splitting strategies, and the results demonstrate that our approach achieves a considerable attack success rate while causing minimal impact on the main task's performance. In this paper, we unveil the inherent security vulnerability of split learning and devote ourselves to fostering the advancement of pertinent defense technologies, making it a valuable contribution to the research community.
\end{abstract}

% }

\section{Introduction}
 \IEEEPARstart{W}ith the rapid development of the Internet of Things, 5G technology and cloud computing technologies, the sheer volume of data has expanded exponentially \cite{ahmed2014survey,yang2017big,sandhu2021big,ZhouFYYWZ20}. Unfortunately, the explosive growth in data has not propelled the development of the deep learning as rapidly as anticipated, constrained by various privacy regulations and laws, such as GDPR \cite{regulation2018general}, CCPA \cite{illman2019california}, and HIPAA \cite{annas2003hipaa}. Distributed learning emerges as a novel training paradigm, leveraging local training and model interactions to bypass centralized data collection, significantly mitigating data privacy concerns, with federated learning \cite{McMahan2017Communication} being its representative. 

\begin{figure}[h]
\centering
\includegraphics[width=8cm]{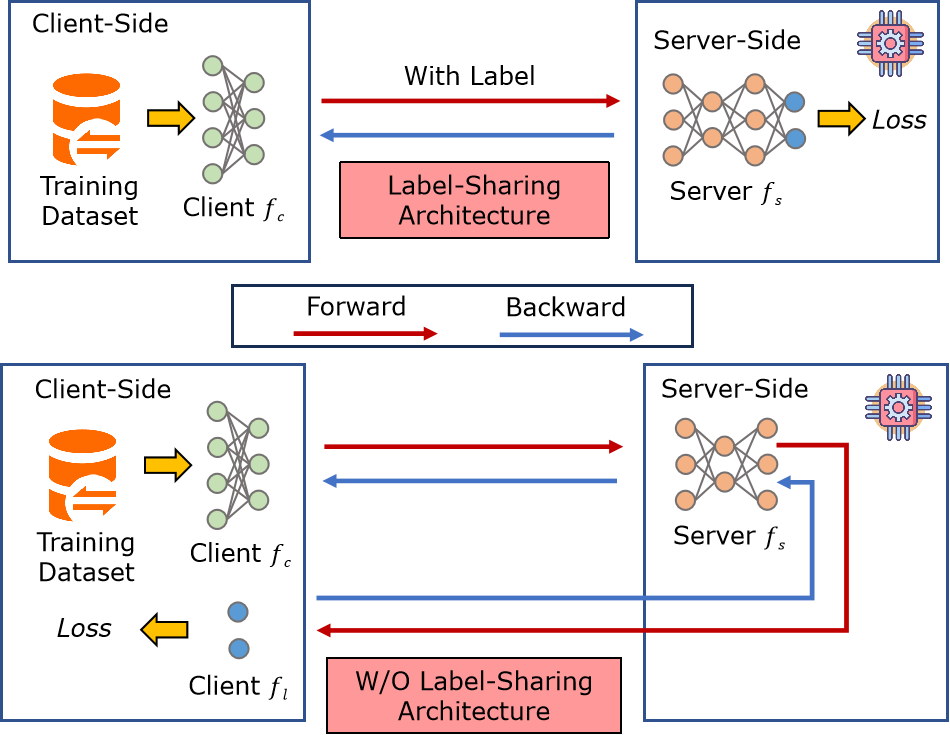}
\caption{Label-sharing and without-label-sharing split learning architectures.}
\label{fig:architecture}
\end{figure}
 
 Although local training can indeed ensure privacy, it comes with substantial computational overhead, which intensifies as the complexity of the model increases. This poses a significant challenge for clients with limited computational resources, making the process less favorable for them. To address this challenge, the concept of split learning has been introduced as a solution. In the split learning paradigm, the client can offload most of the model training computational overhead to the server and prevent the leakage of raw data privacy by uploading the intermediate results \cite{liu2022wireless,thapa2022splitfed,turina2020combining}. Depending on the segmentation pattern of training data, split learning can be divided into label-sharing and without-label-sharing architecture, as illustrated in Fig. \ref{fig:architecture}. In the former, labels are sent to the server, which computes the loss and returns the gradients. In the latter, labels are retained locally, with users calculating the loss themselves. The specific details of both architectures are provided in related work. As the name suggests, the without-label-sharing architecture can more effectively reduce the risk of privacy leakage by preventing the server from obtaining label information.

%Hence, distributed deep learning frameworks (e.g., federated learning and split learning) are proposed \cite{vepakomma2018no,palanisamy2020spliteasy,li2021survey,yang2019federated,lim2020federated,lyu2022privacy,thapa2022splitfed,erdogan2022splitguard,liu2022wireless} to train a model without compromising raw data privacy. In the federated learning paradigm, the clients that participate in the current federated task train a model based on their local data and transmit the parameters or gradient updates to a central server. The server then performs aggregation operations(e.g., taking their average) and redistributes the aggregated model \cite{pillutla2022robust,so2021turbo,zhang2023fedala,rathee2023elsa}. 

%In the split learning paradigm, the client can offload most of the model training computational overhead to the server and prevent the leakage of raw data privacy by uploading the intermediate results \cite{liu2022wireless,thapa2022splitfed,turina2020combining}. Note that this paper just focuses specifically on split learning paradigm.
%the client just computes the feature embedding based on the local data and the first few layers of model and offloads most of the computational overhead to the server-side. The server performs computation tasks based on the rest of the model
Current research predominantly focuses on the privacy issues of split learning, such as data reconstruction \cite{erdougan2022unsplit,pasquini2021unleashing} and label inference \cite{li2021label,kariyappa2023exploit,liu2022clustering,ZhouGFCD0X023}. However, there is a notable lack of research on the security aspects of split learning, such as backdoor attacks, which can cause serious consequences in real-world applications. Generally, a backdoor attack approach aims to manipulate the model in a way that it demonstrates predetermined abnormal behavior when exposed to input data containing a specific trigger, while maintaining accurate predictions for clean data \cite{li2021invisible,qi2022towards,li2022backdoor,saha2020hidden}. For example, in a facial recognition system compromised by a backdoor attack, an attacker could deceive the model using a specific triggered sample to facilitate illegal activities \cite{wuuniid, xue2021backdoors}. Therefore, upon recognizing this issue, we aim to explore the backdoor vulnerabilities inherent in split learning under the without-label-sharing architecture, as such mode offers enhanced security with less knowledge accessible to attackers. In this context,  we ask the following research questions:

\begin{mdframed}[backgroundcolor=black!10,rightline=false,leftline=false,topline=false,bottomline=false,roundcorner=2mm]
\textit{Is it feasible to implement a backdoor attack in the without-label-sharing split learning? If so, how can the stealthiness and effectiveness of such backdoor attacks be ensured?}
\end{mdframed}

Therefore, in this paper, we innovatively investigate the vulnerability of the without-label-sharing split learning architecture against backdoor attacks. Attackers could inject the backdoor triggers in the model while assisting clients in training tasks, so as to  compromise the model for benefits in the future. This attack may cause huge security threats and economic loss in the real world applications, such as IoT devices or financial institutions. Although the backdoor attacks implemented on the server-side are feasible, it is more challenging for the attacker to backdoor the model in split learning due to its lack of access to training data and its label compared with the traditional backdoor attack setting. 

To address the challenges posed by backdoor attacks in without-label-sharing split learning, we have identified the following research questions (RQs):

\begin{itemize}
\item \textbf{RQ1: How can backdoor injection be achieved with limited knowledge?}

In the without-label-sharing architecture, it is impractical for the server to acquire and manipulate training data and corresponding labels, significantly increasing the difficulty of implanting a backdoor.
\item \textbf{RQ2: How can the stealthiness of backdoor attacks be ensured?} 

Considering that the client may employ detection methods to verify the reliability of the gradients during the training process, the attacker must make minor and even no manipulation to the gradients.
\item \textbf{RQ3: How can the effectiveness of backdoor attacks be ensured?} 

It is essential to maintain a high accuracy for the primary task while ensuring a high attack success rate for the backdoor attack.
\end{itemize}

Therefore, in this paper, we propose a novel stealthy backdoor attack (namely \texttt{Dullahan}), which is the first backdoor attack in the without-label-sharing split learning architecture from the server-side perspective, to our best knowledge. 
Different from the previous works that introduce a backdoor by poisoning the training data, \texttt{Dullahan} lies in directly injecting trigger embedding into the server network. For RQ1, we employ trigger embedding and target embedding anchor to replace the training data and corresponding label, respectively. For RQ2, we design a surrogate model building method without modifying the intermediate parameters during standard training to fabricate a more stealthy backdoor attack. For RQ3, we fine-tune the server network by using the collected intermediate data of the training process. Note that our \texttt{Dullahan}, which does not require modifying intermediate data during training, demonstrates robust and satisfactory performance in the presence of attackers, regardless of their knowledge about the client network's architecture.

Our main contributions are summarized as follows: 
\begin{itemize}
\item To the best of our knowledge, this is the first study to reveal the potential backdoor attack risks for the without-label-sharing split learning architecture. We proposed a stealthy and practical backdoor attack strategy from a server-side perspective and achieve a satisfactory performance.

\item To conduct a more stealthy backdoor attack, we propose to inject the backdoor into the server network directly. Specifically, we design a trigger embedding selection method and a target embedding anchor selection approach instead of poisoning the training data and corresponding labels as the traditional backdoor attacks.

\item We conduct extensive experiments of our proposed \texttt{Dullahan} on $3$ datasets and $3$ models. The experimental results demonstrate that our approach yields a fine attack success rate without significantly compromising the primary task's performance.
\end{itemize}

The paper proceeds as follows. Section II reviews the relevant works. Then, we show the threat model in Section III. In Section IV, we present the design of the proposed \texttt{Dullahan}. The experimental results are shown and analyzed in Section V. Finally, we summarize the paper in Section VI.

\section{Related Works}
Split learning substantially diminishes computational expenditures for clients by enabling the sharing of deep learning models with servers, which in turn assume the majority of the training-related computational burden. However, the security implications of split learning have garnered considerable attention from both academic and industrial spheres, leading to the proposal of various attack and defense mechanisms \cite{lee2021triad,dougherty2023stealthy,yu2023backdoor,tajalli2023feasibility,bai2023villain,qiu2022hijack}. In this section, we mainly review the split learning and the relevant works about backdoor attacks and concerning security challenges in split learning.

\subsection{Split Learning}
Split learning is a collaborative machine learning paradigm in which the training process is distributed across multiple devices or servers, allowing data privacy and efficient resource utilization \cite{liu2022wireless,thapa2022splitfed,turina2020combining}. In this setup, each device or server processes only a portion of the data, and the intermediate results are shared with other devices or servers to complete the training process.

Split learning can be divided into two architectures: label-sharing and without-label-sharing, as depicted in Fig. 1. In the label-sharing architecture, the deep learning model is split into two subnetworks: a client network comprising the initial layers and a server network encompassing the remaining layers. They are deployed on the client-side and server-side, respectively. Based on the raw data and client network, the client computes intermediate results (termed smashed data) and uploads these results and their corresponding labels to the server. Then, the server calculates the loss value and transmits the gradients back to the client. In the without-label-sharing architecture, the deep learning model is partitioned into three subnetworks: a client network consisting of the initial layers, a server network comprising the most intermediate layers, and a last network containing the last few layers. The client retains the client network and last network, while the server manages the server network and undertakes the majority of the model training computational overhead. During the model training, the client uploads the smashed data to the server. Then, based on the server network and the received smashed data, the server computes the results and transmits them to the client. Finally, the client calculates the loss value and backward propagates gradients to the server. The server also engages in backward propagation to update the client network. Compared with the label-sharing architecture, the data owner can prevent both raw data and labels from revealing in the without-label-sharing architecture.

\textbf{Evidently, without-label-sharing split learning offers higher privacy and security capabilities because the server receives more limited knowledge. Therefore, this paper primarily considers backdoor attacks under the without-label-sharing split learning architecture, which is more challenging to launch attacks in the label-sharing scenario and thus holds greater practical value.}

\subsection{Backdoor Attack}
Most of the existing backdoor attack methods are designed by poisoning the training data and labels with different strategies. BadNets\cite{gu2019badnets} used fixed corner white blocks as triggers to conduct a backdoor attack. This work first revealed that the deep learning models may be backdoored by poisoning training data. To improve the stealth of the triggers, some invisible backdoor attacks are proposed \cite{liao2018backdoor,li2020invisible,chen2020invisible,saha2020hidden}. Blended \cite{chen2017targeted} proposed to generate the poisoned images by blending the trigger and clean samples in a weighted way. Refool\cite{liu2020reflection} injected reflections as a backdoor into a victim model by using the mathematical modeling of physical reflection models. Considering that all poisoned samples contained the same trigger pattern may be detected easily, some sample-specific backdoor attacks have been proposed. These attacks injected unique triggers for different samples by employing different techniques. Wanet\cite{nguyen2021wanet} fabricated the trigger by using a warping function to improve the stealthiness. SSBA\cite{li2021invisible} generated sample-specific invisible additive noises as the backdoor triggers by using an encoder-decoder network. LF\cite{zeng2021rethinking} proposed to create a smooth backdoor trigger without high-frequency artifacts. Most of the above backdoor attacks try to fabricate poisoned samples that are similar to the clean samples. However, the source label is usually different from the target label. Accordingly, these backdoor attacks can still be detected easily by examining the sample-label relationship of training data. To improve the backdoor attack stealthiness, some clean-label attacks are also proposed \cite{barni2019new,saha2020hidden}. Turner et al.\cite{turner2019label} first create a a backdoor attack method without poisoning the labels. This method just modified some clean samples to conduct an invisible backdoor attack by using adversarial perturbations. Zhao et al. \cite{zhao2020clean} proposed to utilize a universal perturbation trigger instead of a given one as the backdoor trigger. In addition to the aforementioned backdoor attacks, some semantic backdoor attacks \cite{bagdasaryan2021blind,bagdasaryan2020backdoor} and physical backdoor attacks \cite{wenger2021backdoor, li2021backdoor} are also proposed. These attempts improve the practicability of backdoor attacks in real-world applications.

\subsection{Split Learning Security}
Many researchers pay mainly attention to the privacy problem of split learning \cite{liu2022clustering,erdougan2022unsplit,pasquini2021unleashing,li2021label}. There are only a few researches on the security issues of split learning. Fan et al. \cite{fan2023robustness} introduced a two-stage attack strategy involving the training of a surrogate model and the generation of adversarial examples, which has demonstrated a high success rate and exposed the susceptibility of split learning to adversarial attacks. Bai et al. \cite{bai2023villain} proposed a backdoor attack framework named VILLAIN for vertical split learning, a backdoor attack that enabled attackers to achieve high inference accuracy for targeted label samples. Tajalli et al. \cite{tajalli2023feasibility} proposed two attack methods for split learning in the label-sharing architecture: one utilizing a surrogate client and the other employing an auto-encoder to corrupt the model. Their experiments performed poorly against backdoor attacks. The authors supposed that split learning exhibited robustness against backdoor attacks when the server's involvement is limited to its training capabilities without access to or manipulation of client-side data. Following this, Yu et al. \cite{yu2023backdoor} proposed two backdoor attack frameworks from both the server and the client perspectives. For client-side attackers, they can perform backdoor attacks by inserting backdoor samples into the training data. For server-side attackers, they leveraged the server's control over the training process to shape the optimization direction of the model. Both the proposed attacks can achieve high attack accuracy without reducing the performance of the main task. However, the proposed attacks require gradient modifications, which can be easily detected by some abnormal gradient detection methods, thus lacking stealth. Moreover, their proposed backdoor attacks are only suitable for split learning in a label-sharing architecture. 

\textbf{Compared with these existing backdoor attacks by poisoning training data, our \texttt{Dullahan} eliminates the need to manipulate training data or labels and avoids the backpropagation gradient alterations during training, making it less detectable and posing a greater threat to the split learning paradigm.}

\section{Threat Model}
In this section, we define the threat model of the proposed \texttt{Dullahan} for the without-label-sharing split learning architecture. The server, acting as the attacker, employs a four-step backdoor attack: $\textcircled{1}$ Surrogate Model Building $\rightarrow$ $\textcircled{2}$ Trigger Embedding Selection $\rightarrow$ $\textcircled{3}$ Target Label Selection $\rightarrow$ $\textcircled{4}$ Backdoor Injection. The detailed scheme will be described in Section \ref{scheme}, and the specific threat model is illustrated in the Fig. The attacker's goal, knowledge, and capability are outlined as follows:

\begin{figure}[h]
\centering
\includegraphics[width=8cm]{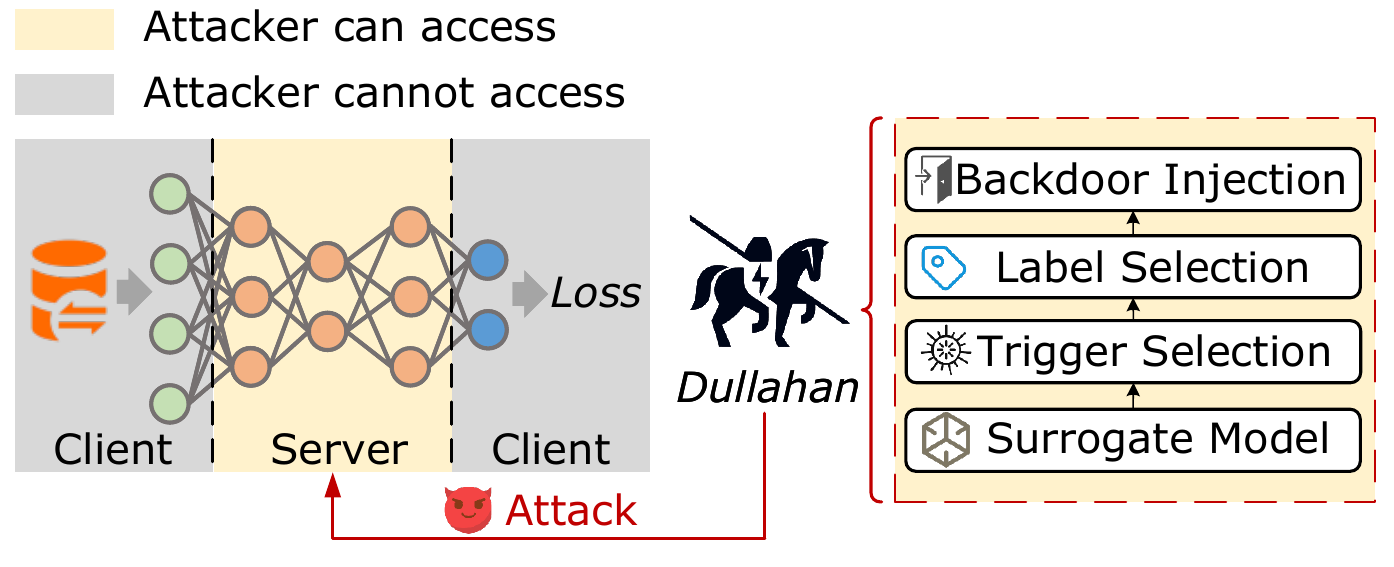}
\caption{Threat model of backdoor attack in  without-label-sharing split learning.}
\label{fig:threatmodel}
\end{figure}

\begin{itemize}
    \item \textbf{Attacker's goal}: The attacker (e.g., the server provider) aims to inject a backdoor into the model when assisting the client in training a model. The model correctly predicts the label for benign samples and outputs the targeted label for triggered samples during the inference phase. Note that we just consider the targeted backdoor attack, which is a special case of untargeted backdoor attacks where the triggered sample can induce the model to output any wrong label. Compared with untargeted backdoor attacks, targeted backdoor attacks are inherently more challenging.
    
    \item \textbf{Attacker's knowledge}: We assume that the attacker has enough computational resources and access to a public auxiliary dataset $X_{\rm{aux}}$ with the same domain as the client's training dataset, which is the same as the assumption in the \cite{pasquini2021unleashing,tajalli2023feasibility,yu2023backdoor}. For example, if the model is trained on the digital images, the $X_{\rm{aux}}$ is also composed of the digital images. The amount of $X_{\rm{aux}}$ is just nearly 10\% of the training dataset, with no overlap assumed between the two. Besides, we consider two situations about the model knowledge of the attacker. The first is that the attacker is aware of the architecture of the client network $f_c$ but has no information about the weights of $f_c$. We argue that we usually train a model based on some popular model architectures so the server-side can infer $f_c$'s architecture based on the server network $f_s$. Another one is that the attacker has no information about the client network. Compared with the existing assumption, our threat model is more pragmatic than assumptions in related works \cite{vepakomma2019reducing, vepakomma2018no}, where the adversary is assumed to have direct access to leaked pairs of private data and the smashed data.
    
    \item \textbf{Attacker’s capability}: To conduct a stealthy backdoor attack, we impose more strict limitations on the attacker's capabilities compared to existing threat models. In our threat model, the attacker cannot modify both the embedding during forward-propagation and the gradients during back-propagation. This constraint precludes malicious influence during training, ensuring all parameters received by the client are untampered to avoid being detected by potential defense strategies. The attacker only records the smashed data provided by the client and the trained server network $f_s$ during training process. Then, the attacker can train a surrogate model of the client network $f_c$ based on the auxiliary dataset $X_{\rm{aux}}$. Moreover, the attacker can only manipulate and modify the server network's parameters after accomplishing normal model training. 
\end{itemize}

\section{Design of \texttt{Dullahan}} \label{scheme}
In this section, we provide an overview of the proposed \texttt{Dullahan} followed by an in-depth exposition of its design.

\begin{figure*}[htp]
\centering
\includegraphics[width=16cm]{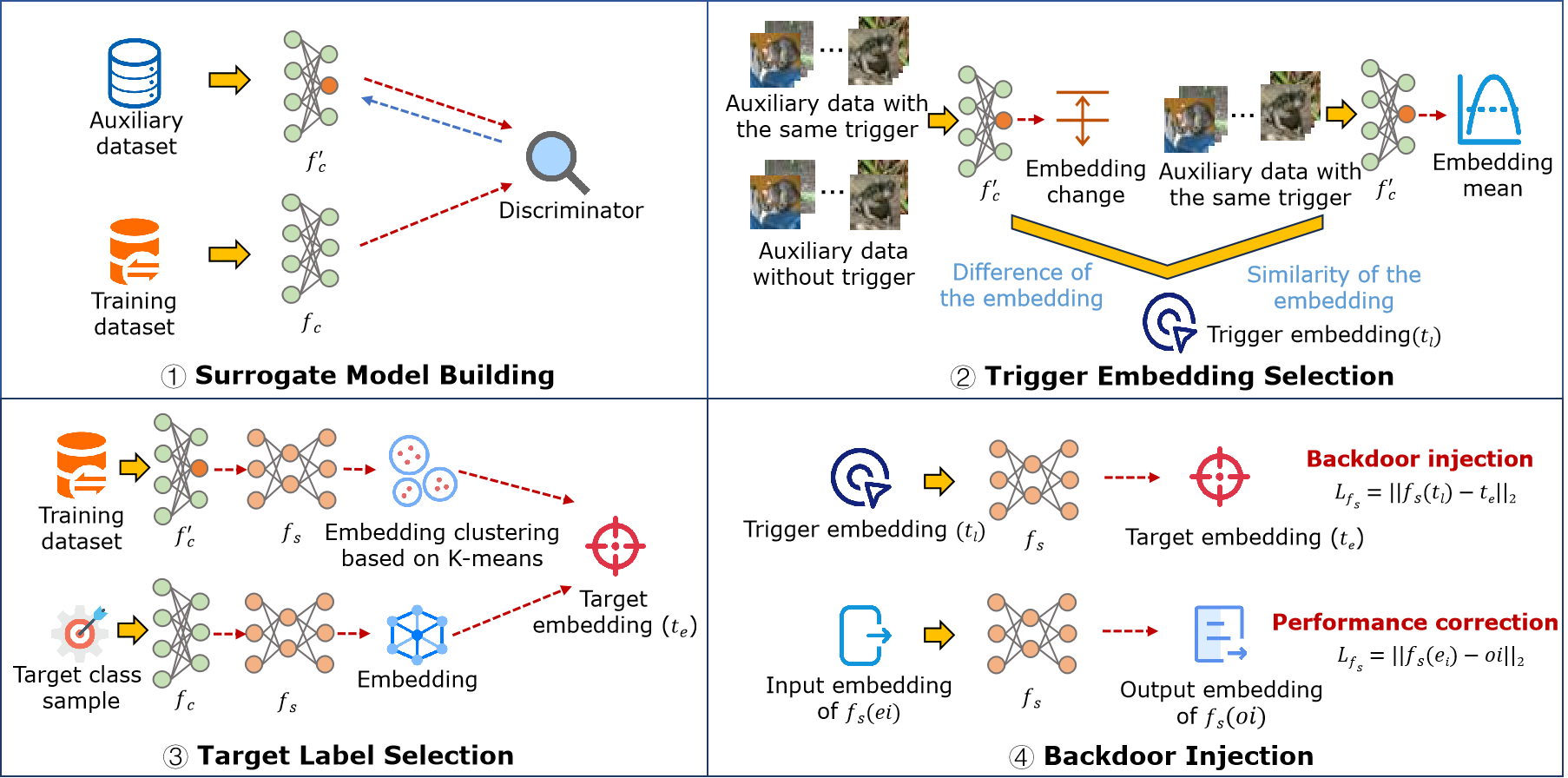}
\caption{The overview of the proposed \texttt{Dullahan}.}
\label{fig:framework}
\end{figure*}

\subsection{Overview of \texttt{Dullahan}}
In this paper, we assess the susceptibility of split learning to backdoor attacks by introducing a method orchestrated by the server in the without-label-sharing split learning architecture. Given the challenges in accessing the training data and corresponding labels, traditional backdoor injection techniques that rely on directly manipulating training data are considered infeasible. Consequently, our approach diverges from conventional methods by embedding a trigger directly into the server network, circumventing the need for training data manipulation. Our \texttt{Dullahan} unfolds through the following stages. Firstly, we build a surrogate model $f'_{c}$ based on the auxiliary dataset to approximate the client network $f_{c}$. Secondly, the trigger embedding is calculated and selected based on the surrogate model $f'_{c}$ by designing a statistical approach that finds out those bits that have the greatest impact on the backdoor trigger. Then, we replace the role of target label with the target embedding anchor, by utilizing a $K$-means cluster algorithm to locate the target embedding anchor of the target label. Finally, we inject the backdoor into the server network $f_{s}$ by computing the loss between the trigger embedding and the target embedding anchor. Moreover, to maintain the performance of the main task, we also fine-tune $f_{s}$ with the intermediate data collected. Fig. \ref{fig:framework} illustrates the procedural framework of our \texttt{Dullahan}.

\subsection{Detailed Design of \texttt{Dullahan}}

In this section, we present the detailed design of our \texttt{Dullahan}. The notations used in this paper are shown in Table \ref{tab:Notions}.

\begin{table}
    \caption{Notions and corresponding definitions.}
    \label{tab:Notions}
    \centering
     \setlength{\tabcolsep}{7mm}{
    \begin{tabular}{cc}
        \hline
        \textbf{Notation}&\textbf{Explanation}\\
        \hline
         $f_{c}$ & Client network \\
         $f_{s}$ & Server network\\
         $f_{l}$ & Last network\\
         $X$ & Training data\\
         $X_{\rm{aux}}$ & Auxiliary dataset\\
         $f'_{c}$ & Surrogate model\\
         $f_{D}$ & Discriminator\\
        \hline
    \end{tabular}}
\end{table}

\subsubsection{Surrogate Model Building}
It's necessary to know the trigger patch before conducting a backdoor attack, which is hardly guaranteed in without-label-sharing split learning architecture, due to the client's full control over the training data and the client network $f_{c}$. Even if the attacker injects a backdoor into the server network, it is impractical for the attacker to obtain the exact pattern of the trigger patch pasted on the input samples due to the lack of a shallow client model. Therefore, we have to train a surrogate model to assist backdoor injection, which is necessary for our attack method. Different from the existing surrogate model building methods \cite{yu2023backdoor, tajalli2023feasibility, pasquini2021unleashing}, which try to make the client network $f_{c}$ get close to the surrogate model $f'_{c}$ by back-propagation forged gradients to $f_{c}$. However, the forged gradients may be detected by some malicious gradient detection methods. To conduct a more stealthy attack, we build a surrogate model $f'_{c}$ by letting the surrogate model approximate the client network $f_{c}$ without manipulating any forward and back-propagation information. 

We assume that the training data and auxiliary dataset are $X$ and $X_{\rm{aux}}=\{x_1,x_2,...,x_n\}$, respectively. The attacker trains two different networks, namely $f'_{c}$ and $f_{D}$. These serve very distinct roles, more precisely:

\begin{itemize}
    \item $f'_{c}$: $f'_{c}$ can be seen as a mapping between a data space $X$ (i.e., where training samples are defined) and a feature space $Z$ (i.e., where the smashed data are defined). Note that $f'_{c}$ can be initialized with the same architecture as $f_{c}$ or a different architecture. Moreover, our goal is that $f'_{c}$ should be as similar as possible to $f_{c}$. That is, $|f'_c(x)|\approx|f_c(x)|$, where $x$ is a sample.
    \item $f_{D}$: $f_{D}$ is a discriminator\cite{goodfellow2014generative} that indirectly guides $f'_{c}$ to learn a mapping between the private data and the feature space defined from the $f_{c}$.
\end{itemize}

To be specific, the discriminator $f_{D}$ \cite{goodfellow2014generative} is trained to distinguish between the feature space induced from $f'_{c}$ and the one induced from the client network $f_{c}$. $f_{D}$ takes $f_{c}(X)$ (i.e., the smashed data) and $f'_{c}(X_{\rm{aux}})$ as input and is trained to assign high probability to the former and low probability to the latter. That is, owing to the $X_{\rm{aux}}$ with the same domain as the training dataset $X$, we can make the output distribution of $f'_{c}$ close to that of $f_{c}$. More formally, at each training iteration (i.e., when the client sends smashed data to the server), the weights of $f_{D}$ are tuned by minimizing the following loss function:
\begin{equation} \label{equation1}
    \mathcal{L}_{f_{D}}=log(1-f_{D}(f_{c}(X)))+log(f_{D}(f'_{c}(X_{\rm{aux}}))).
\end{equation}
After each local training step for $f_{D}$, to make $f'_{c}(X_{\rm{aux}})$ as close as possible to $f_{c}(X)$, we tune the weights of $f'_{c}$ by minimizing the following loss function:
\begin{equation} \label{equation22}
    \mathcal{L}_{f'_{c}}=log(1-f_{D}(f'_{c}(X_{\rm{aux}}))).
\end{equation}
By optimizing the above Equation (\ref{equation1}) and (\ref{equation22}), we can obtain a surrogate model $f'_{c}$ which is similar to the client network $f_{c}$.

\begin{figure*}[htp]
\centering
\includegraphics[width=15cm]{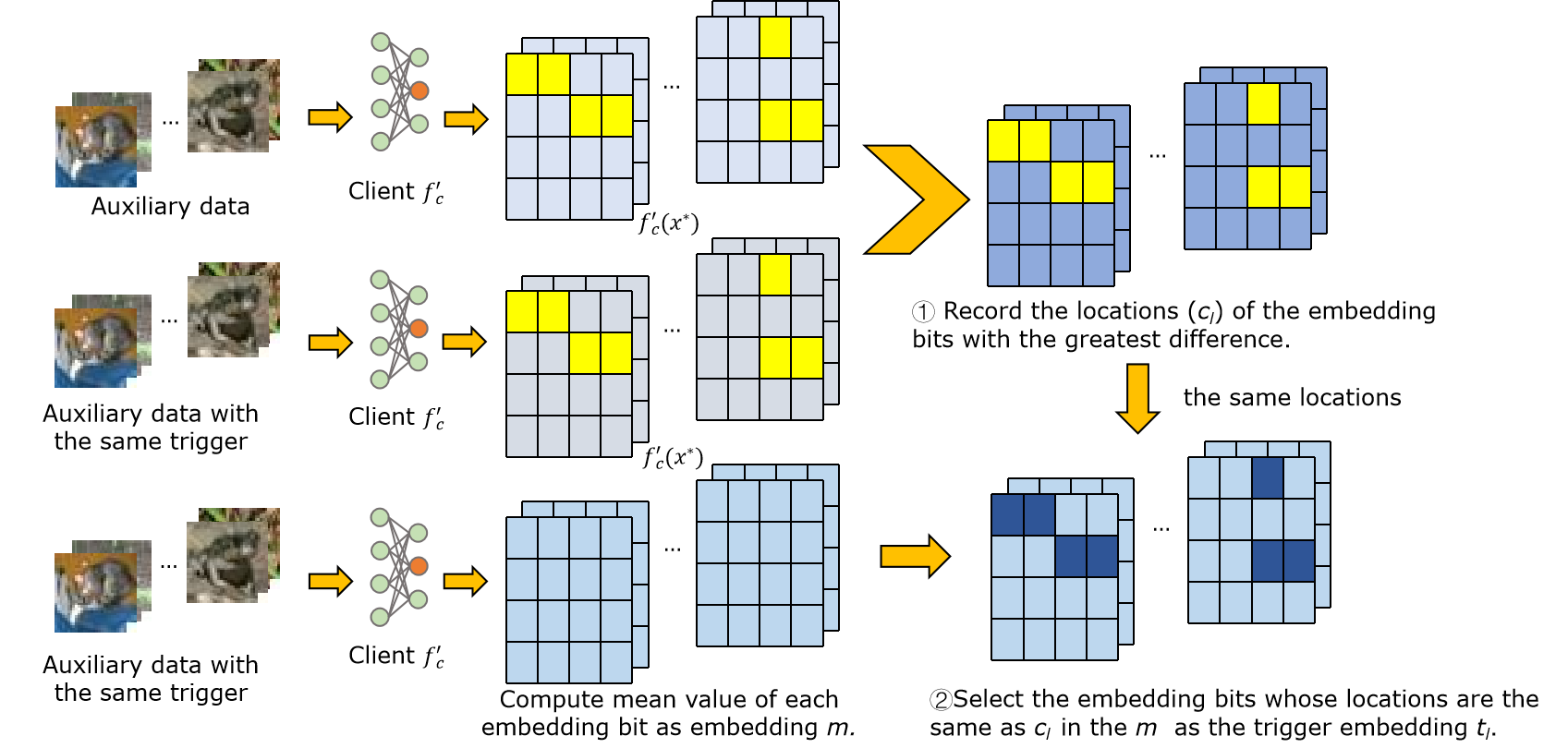}
\caption{The overview of trigger embedding selection.}
\label{fig:Trigger Embedding Selection}
\end{figure*}

\subsubsection{Trigger Embedding Selection}

Even if we have obtained the surrogate model $f'_{c}$, it is still difficult for us to inject the backdoor as the conventional methods due to the lack of control over the training data. Therefore, we plan to conduct a backdoor attack by directly injecting a backdoor trigger embedding into the server network $f_{s}$. However, choosing an available trigger embedding is a vital issue. 

Hence, we propose a statistical method to select the trigger embedding. The main idea of this method is to locate the embedding bits where the trigger has the most impact by finding the difference between the clean samples and the samples with the a trigger after inputting the surrogate model. Fig. \ref{fig:Trigger Embedding Selection} shows the workflow of trigger embedding selection. There are two phases shown as follows:

(1) To explore the difference of embeddings between the clean sample and the backdoored sample after inputting the client network $f_{c}$, we add the a trigger to the auxiliary dataset $X_{\rm{aux}}$ to obtain the backdoor dataset $X'_{aux}=\{x^*_1,x^*_2,...,x^*_n\}$ at first. Then, we compute the mean distance (denoted as $c$) of each embedding bit between the clean sample and the backdoored sample by employing the surrogate model $f'_{c}$, which is denoted as:
\begin{equation} \label{equation3}
    c=\frac{1}{n}\sum_{i=1}^n|f'_{c}(x_i)-f'_{c}(x^*_i)|.
\end{equation}
where $n$ is the sample number of auxiliary dataset $X_{\rm{aux}}$, $x_i$ is a clean sample and $x^*_i$ is the clean sample with a trigger. Then, we select and record a certain number (e.g., $50$) of bits' location (denoted as $c_l$) of the embedding with the greatest distance. The larger the distance indicates that these embedding bits may be more influenced by the backdoor trigger. 

(2) After finding out the location of the embedding bits affected most by the backdoor trigger patch, we further determine what the specific values of these most influential bits are. Therefore, we try to achieve it by computing the mean value (denoted as $m$) of the embedding of the samples with the same trigger, which is denoted as:
\begin{equation} \label{equation4}
    m=\frac{1}{n}\sum_{i=1}^nf'_{c}(x^*_i).
\end{equation}
Finally, the embedding bits whose locations are the same as $c_l$ in the $m$ are selected as the trigger embedding $t_l$. 

\subsubsection{Target Embedding Anchor Selection}
Since the server-side attacker can only gain the input and output embeddings of the server network in the without-label-sharing split learning architecture, it is difficult for the server-side to obtain and manipulate the label. Notably, it is necessary to control the target label when conducting a backdoor attack. Therefore, we propose to employ a fixed target embedding anchor to replace the target label. Nonetheless, there is still a great challenge in distinguishing which label the target embedding anchor belongs to. To address this challenge, we propose a label inference method based on $K$-means clustering algorithm. The workflow of this method is shown in Fig. \ref{fig:Target Embedding Selection}. It mainly contains the following two phases:

\begin{figure}[htp]
\centering
\includegraphics[width=8cm]{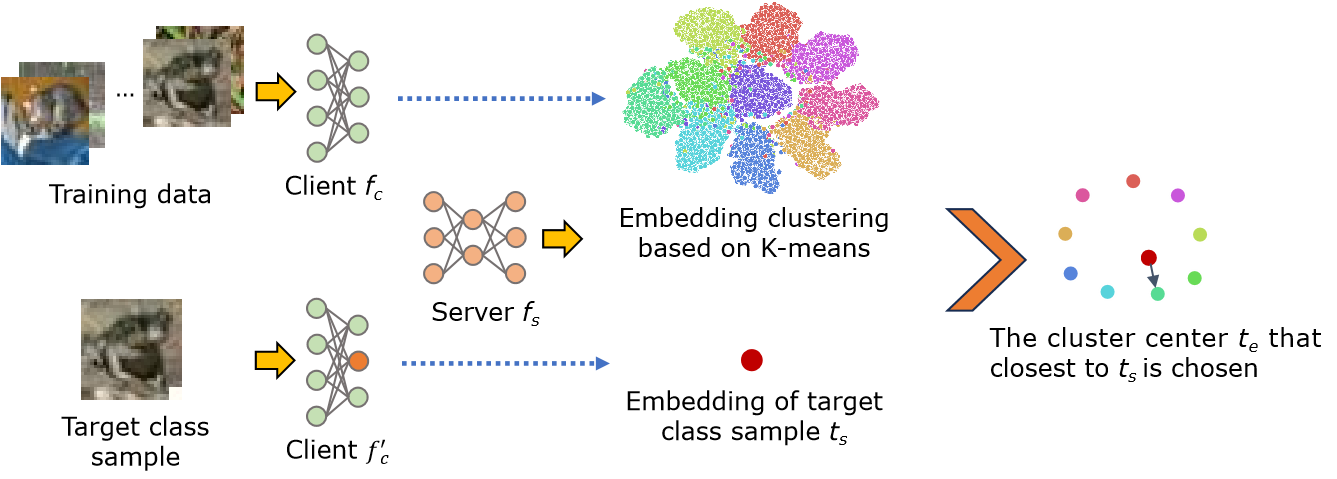}
\caption{The overview of target embedding anchor selection.}
\label{fig:Target Embedding Selection}
\end{figure}

(1) The server-side attacker computes the corresponding outputs ($O=\{o_1, o_2, ..., o_n\}$) based on the smashed data ($E=\{e_1, e_2, ..., e_n\}$) provided by the client and the server network that completes the standard training. Then, the server-side attacker groups the outputs into $K$ clusters with similar characteristics as \cite{liu2022clustering}, where $K$ is the number of classes of the classification task. The objective of clustering is to minimize the within-cluster sum of similarity measurements (e.g., based on the Euclidean distance) and obtain $K$ clusters. Each cluster represents one class. The embeddings of $K$ cluster centers are recorded as $CT=\{CT_1, CT_2,...,CT_K\}$.

(2) The server-side attacker selects all the target class samples ($X_t=\{x_1, x_2, ..., x_m\}$) from the auxiliary dataset $X_{\rm{aux}}$. Then, each target class sample $x_{j}$ is inputted into $f'_{c}$ to obtain the smashed data $f'_{c}(x_{j})$. Then, the $f'_{c}(x_j)$ is inputted into $f_{s}$ to gain the embedding denoted as $t_j=f_{s}(f'_{c}(x_j))$. To further determine which cluster center belongs to the target class, we compute the sum of the distance between each cluster center and $t_j$, respectively, which is denoted as: 
\begin{equation} \label{equation6}
    d_i=\sum_{i=1}^m||CT_i-t_j||_2.
\end{equation}
where $CT_i$ is the $i$-th cluster center. Finally, the nearest cluster center to the target class samples (i.e., the minimum $d_i$) is selected as the fixed target embedding anchor ($t_e$) to assist backdoor injection.

\subsubsection{Backdoor Injection}
After choosing the trigger embedding and the embedding anchor of the target class, we inject the backdoor into the server network ($f_s$) by computing the loss function:
\begin{equation} \label{equation7}
    \mathcal{L}_{f_s}=||f_s(t_l)-t_e||_2.
\end{equation}
To guarantee the performance of the original model, we also fine-tune the server network by using the collected inputs ($E$) and computed corresponding outputs ($O$), just for further optimizing $\mathcal{L}_{f_s}$. It is denoted as:
\begin{equation} \label{equation8}
    \mathcal{L}_{f_s}=||f_s(e_i)-o_i||_2.
\end{equation}
where $e_i$ denotes the embedding from the client during the last epoch training, and $o_i$ represents the corresponding output when $e_i$ is inputted into the normally trained server network.

After optimizing Equation (\ref{equation7}) and (\ref{equation8}) at the same time, the backdoor can be injected into the training model. Note that our \texttt{Dullahan} can be conducted after finishing the normal training process. Compared with the existing backdoor methods \cite{tajalli2023feasibility,yu2023backdoor} that requiring modifying the gradients during backpropagation, our \texttt{Dullahan} just needs to implicitly collect the smashed data provided by the client and furtively calculate the corresponding outputs based on the trained server network. Therefore, our \texttt{Dullahan} is more stealthy. 

\section{Experiments}
In this section, we comprehensively evaluate the performance of our \texttt{Dullahan}. First, we show the experimental settings, including the experimental environment, datasets, and networks. Then, we analyze and summarize the experimental results.   

\subsection{Experimental Settings}

\subsubsection{Experimental Environment}
All experiments were conducted on a server equipped with Intel(R) Xeon(R) Gold 6346 CPU, 3.10GHz processor, 256GB RAM, and NVIDIA GeForce RTX 3090. PyCharm and PyTorch are used to deploy the model and complete other relevant experiments.

\subsubsection{Datasets}
We perform our experiments on three image datasets: MNIST\cite{deng2012mnist}, F-MNIST\cite{xiao2017fashion}, and CIFAR-10\cite{krizhevsky2009learning}. The introduction of the above three datasets is shown as follows.

\textbf{MNIST}:
The MNIST database is a database of handwritten digits, including 60,000 training images and 10,000 testing images from 10 classes.

\textbf{F-MNIST}:
The F-MNIST dataset is a database of fashion images, including 60,000 training images and 10,000 testing images from 10 classes.

\textbf{CIFAR-10}:
The CIFAR-10 dataset is a database of tiny images, including 50,000 training images and 10,000 testing images from 10 classes.

\subsubsection{Networks}
We perform our experiments on three networks, including ResNet50, ResNext50, and VGG16. The description of the above three networks is shown as follows:

\textbf{ResNet50}: 
ResNet50 is a convolutional neural network of 50 layers consisting of residual blocks. 

\textbf{ResNext50}: 
ResNext50 is a derived type of ResNet50, which divides channels into multiple groups.

\textbf{VGG16}: 
VGG16 is a convolutional neural network model based on the Visual Geometry Group model design, with a depth of 16 layers.

\subsubsection{Evaluation Metrics}
To evaluate the performance of the proposed \texttt{Dullahan}, we employ the accuracy of the main task (Baseline) without backdoor injection, the accuracy of the main task (ACC) with backdoor injection, and the attack success rate (ASR). ACC shows the impact of backdoor injection on the main task, and ASR is used to evaluate the performance of the backdoor attack method.

\subsubsection{Relevant Parameter Settings}\label{label:setting}

In order to evaluate the performance of the proposed \texttt{Dullahan} for split learning, we operate extensive experiments on three networks (ResNet50, ResNext50, VGG16) and three datasets (MNIST, F-MNIST, CIFAR-10). Concretely, The original training dataset is divided into two parts: the training dataset and the auxiliary dataset. Each class in the auxiliary dataset has the same number of samples. The number of the relevant datasets is shown in Table \ref{tab:dataset}. We modify the size of $4\times4$ pixels as a white square trigger patch. The size of the chosen trigger embedding is set as $50$ bits by default.  Table \ref{tab:networks} in Appendix shows the networks and split strategies when knowing the client network's architecture.  

Moreover, to inject the backdoor into the $f_s$, we record all the inputs ($Inp_s$) of the $f_s$ in the last epoch during the standard training process. We also compute corresponding outputs($Out_s$) based on $Inp_s$ and the trained server network $f_s$. Then, we split the $Inp_s$ into multiple batches ($bat_i$, $i=\{1,2,...,n\}$, where $n$ represents the number of the batches). For each batch, we add the trigger embedding to the $bat_i$ as the triggered embedding ($Bbat_i$). The above $bat_i$ and $Bbat_i$ are combined into one new batch. Further, the new batch is fed into the $f_s$ just to finish the backdoor injection. Besides, to avoid occasionality, we record and employ the mean ACC and ASR for executing $6$ epochs to $10$ epochs backdoor injection operations as the final results.

\begin{table}
    \centering
    \caption{The number of the relevant datasets.}
    \label{tab:dataset}
    \begin{tabular}{cccc}
        \toprule
        Datasets & Training Dataset & Test Dataset & Auxiliary Dataset\\
        \midrule
        MNIST & 54000 & 10000 & 6000\\
        F-MNIST & 54000 & 10000 & 6000\\
        CIFAR-10 & 45000 & 10000 & 5000\\
       \bottomrule
    \end{tabular}
  
\end{table}

\begin{figure}[htp]
\centering
\includegraphics[width=8cm]{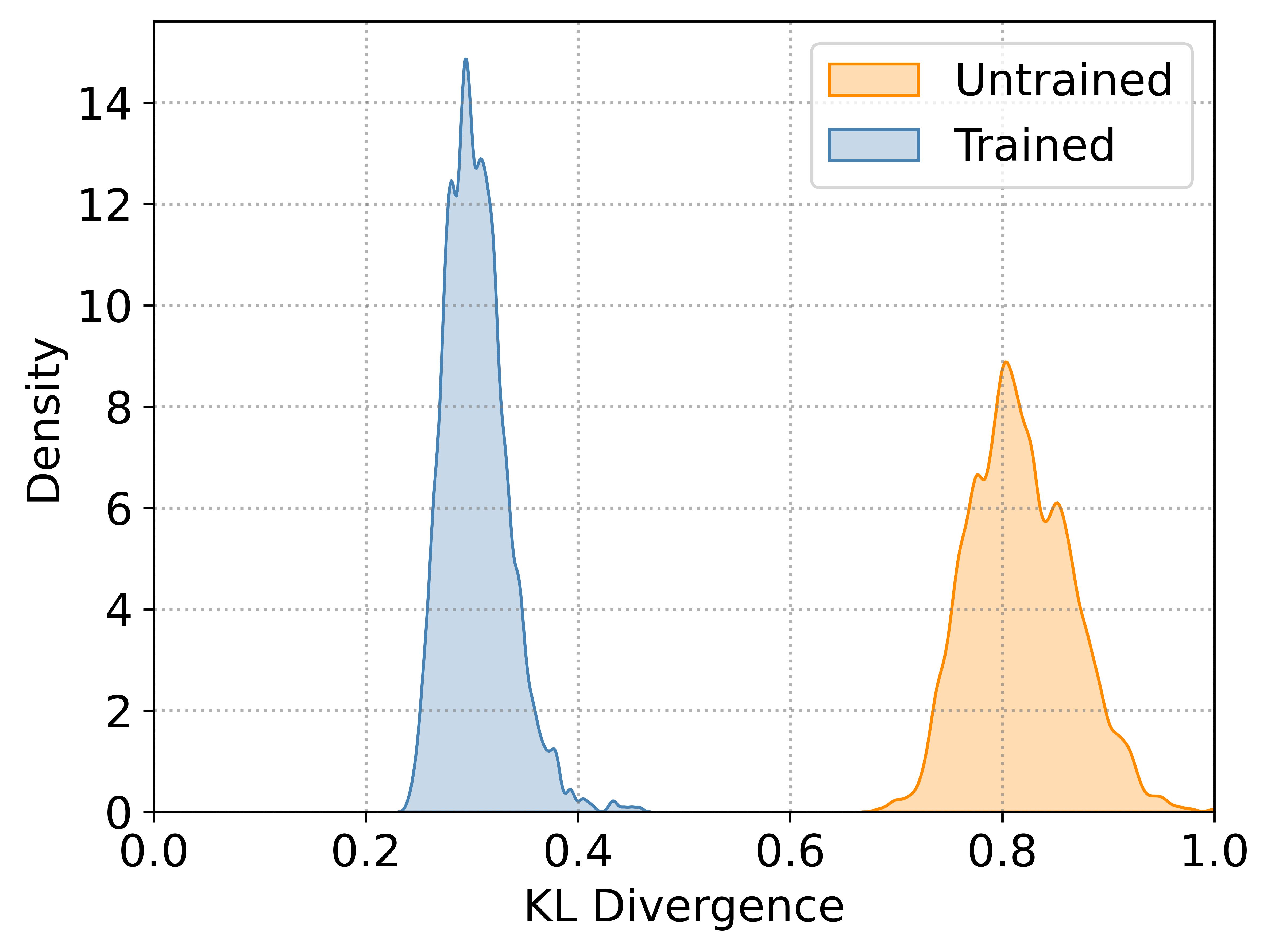}
\caption{The performance of surrogate model.}
\label{fig:Surrogate}
\end{figure}

\subsection{Evaluation of Surrogate Model}
In the proposed \texttt{Dullahan}, the surrogate model plays a critical role in backdoor injection. Hence, in this section, we evaluate the similarity between the surrogate model and the client network. Here, we take an example of the attacker knowing the architecture of the client network on ResNet50. We train the surrogate model nearly 200 epochs by using the auxiliary dataset and the collected intermediate data. We can train the surrogate model after finishing the training process and just require recording all the inputs in the last epoch of standard training. We employ KL divergence of the output from the client network and that from the surrogate model to evaluate the surrogate model's performance. That is, for the same input sample, the embedding outputs from the client network and the surrogate model are more similar. This indicates that the performance of the surrogate model is better. Fig. \ref{fig:Surrogate} illustrates the KL Divergence values for between the client network and the initial surrogate model (orange curve), and between the client network and the post-optimized surrogate model (blue curve). We know that the surrogate model is more similar to the client network after our optimizing mechanism. More intuitively, we also employ ACC to show the effectiveness of our surrogate model building method. We combine the surrogate model, the trained server network, and the last network to form a complete model. For the untrained complete model, the ACC of the main task is $10.12\%$. After optimizing the surrogate model, the ACC of the complete model is $55.48\%$. Compared with the ACC ($81.30\%$) of the original model, we know that there are still some differences between the client network and the surrogate model. However, it is enough for us to conduct a backdoor attack.

\subsection{Selection of Target Embedding Anchor}
\begin{figure}[htp]
\centering
\includegraphics[width=8cm]{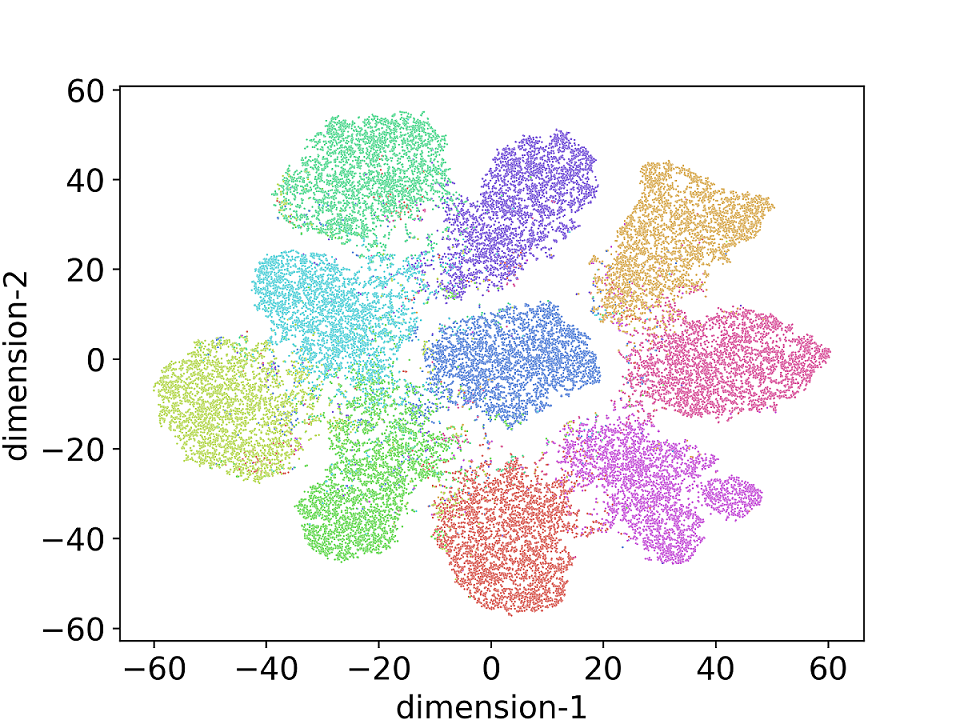}
\caption{The performance of embedding clustering.}
\label{fig:embedding_clustering}
\end{figure}

\begin{figure}[htp]
\centering
\includegraphics[width=8cm]{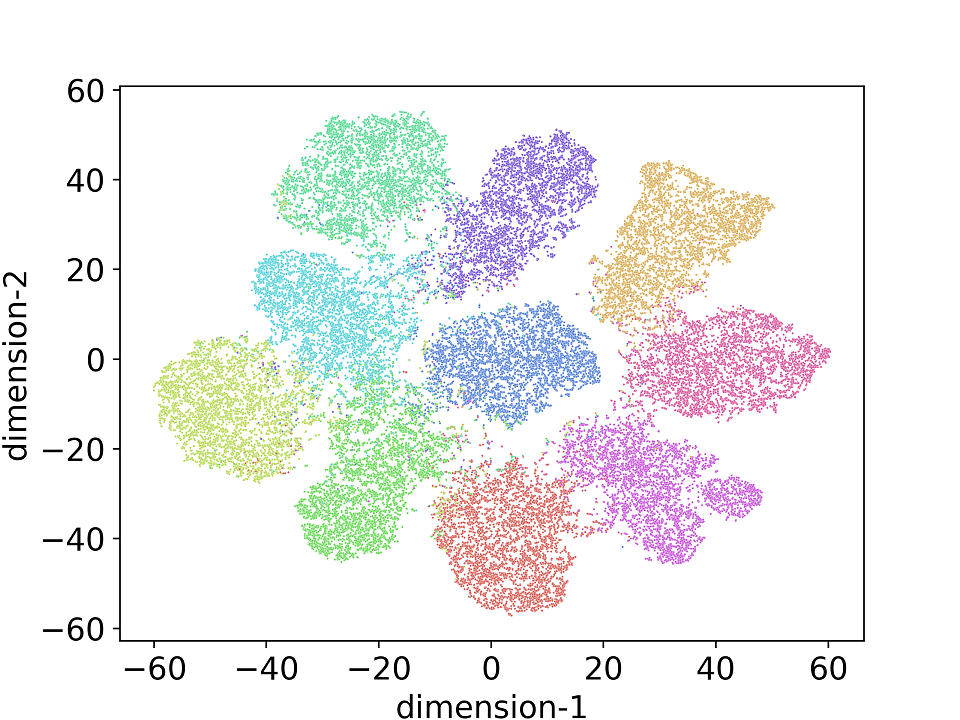}
\caption{The performance of true label clustering.}
\label{fig:label_clustering}
\end{figure}

Since the attacker cannot directly obtain and manipulate the sample labels, we propose to use the target embedding anchor to replace the sample label. The intuition is that the clustering results of the output embedding from the server network are similar to the true label distribution. Fig. \ref{fig:embedding_clustering} shows the $K$-means clustering results of the embedding. Fig. \ref{fig:label_clustering} reveals the clustering results that we mark the true label on the embedding outputted from the server network. From Fig. \ref{fig:embedding_clustering} and Fig. \ref{fig:label_clustering}, we can know that the clustering result of the embedding is extremely similar to that of the true label. It indicates that it is feasible to use the target embedding anchor to replace the true label to conduct a backdoor attack.

\subsection{Evaluation of Proposed Dullahan}
In this section, we present the experimental results of our proposed Dullahan from various perspectives.
\subsubsection{\textbf{Backdoor Attack Performance}}

The backdoor injection experimental results are shown in Table \ref{tab:main result}. From Table \ref{tab:main result}, we know that our \texttt{Dullahan} can achieve a good attack performance and have little influence on the main task performance. The highest ASR is $100.00\%$, and the lowest ASR is more than $75.00\%$. Moreover, the maximum accuracy loss is within $1.18\%$. For example, for ResNet50 on MNIST, the ACC is $99.26\%$. Compared with the original main task accuracy baseline of $99.2\%$, the accuracy increases $0.06\%$. Moreover, the ASR is almost $94.20\%$.

\begin{table}[h]
    \caption{The performance (\%) of the proposed \texttt{Dullahan}.}
    \label{tab:main result}
    \centering
    \begin{tabular}{ccccc}
        \midrule
         Model & Dataset & Baseline & ACC & ASR\\
        \midrule
        \multirow{3}{*}{ResNet50} & MNIST & 99.20 & 99.26 & 94.20\\
        & F-MNIST & 91.60 & 91.62 & 90.40\\
        & CIFAR-10 & 81.30 & 80.18 & 75.40\\
        \midrule
        \multirow{3}{*}{ResNext50} & MNIST & 99.20 & 99.20 & 76.00\\
        & F-MNIST & 90.50 & 90.48 & 81.30\\
        & CIFAR-10 & 83.00 & 81.95 & 79.40\\
        \midrule
        \multirow{3}{*}{VGG16} & MNIST & 99.10 & 99.02 & 100.00\\
        & F-MNIST & 92.40 & 91.97 & 98.30\\
        & CIFAR-10 & 83.80 & 82.62 & 81.40\\
        \midrule
    \end{tabular}
\end{table}

Further, we study and analyze various factors (including the size of trigger embedding and the split strategy) that may affect the performance of the proposed \texttt{Dullahan}.
The relevant experimental results are shown as follows.

\subsubsection{\textbf{Selection of Trigger Embedding} }

\begin{figure*}[htp]
\centering
\includegraphics[width=18cm]{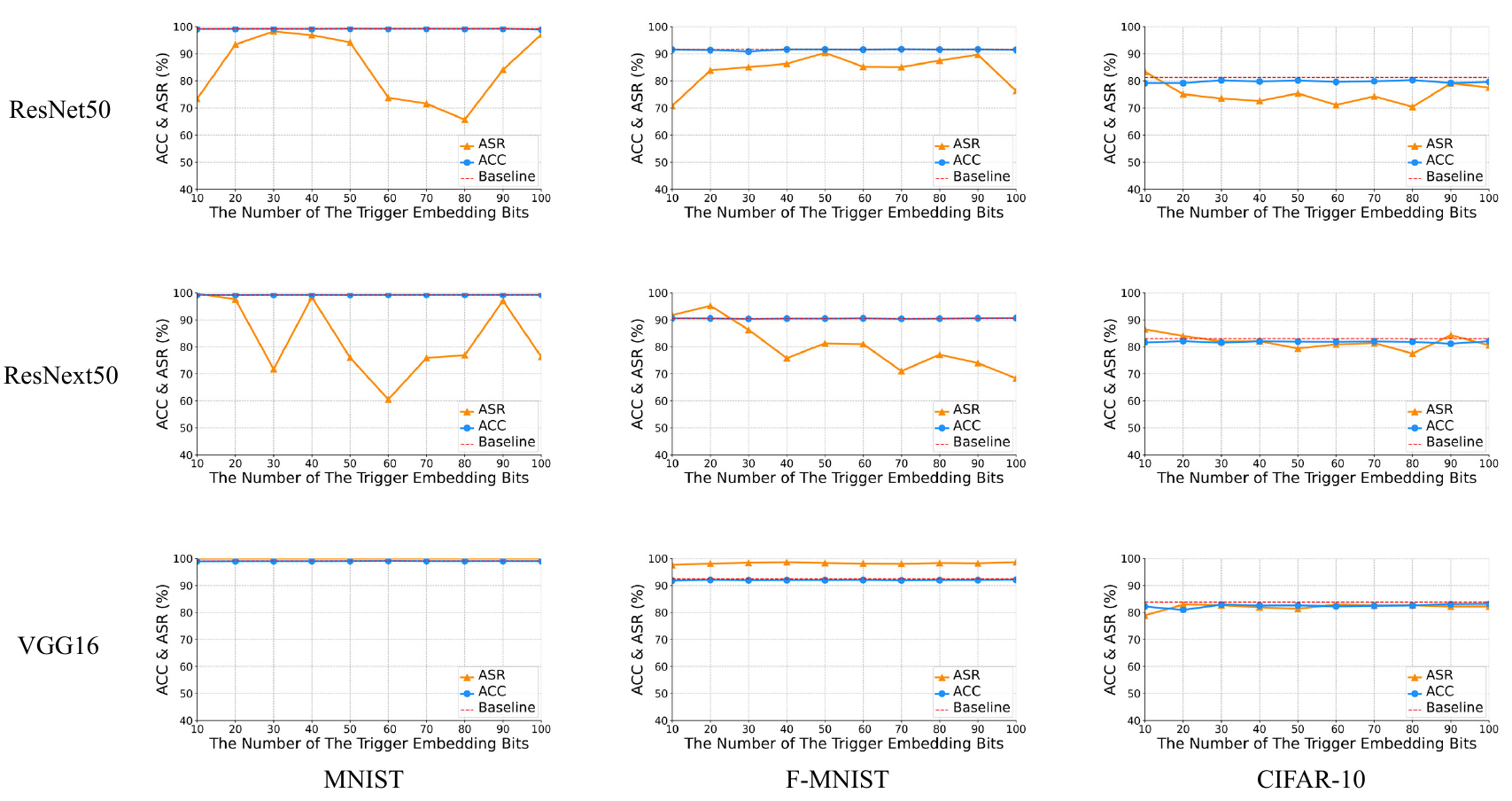}
\caption{The attack performance on three models and three datasets with the different number of trigger embedding bits. The figures from top to bottom row are the experimental results on ResNet50, ResNext50, and VGG16, respectively. The figures from left to right column are the experimental results on MNIST, F-MINST, and CIFAR-10, respectively.}
\label{fig:Trigger_Embedding_Results}
\end{figure*}

In our \texttt{Dullahan}, we try to inject a backdoor into the server network by inserting a trigger embedding. However, the trigger embedding is chosen by analyzing the influence of the trigger on the embedding bits. It is difficult to determine the exact number of the critical embedding bits, which implies that the chosen trigger embedding may contain some unimportant bits that don't matter to the trigger patch. Therefore, the number of the chosen trigger embedding bits may influence the attack performance. We conduct experiments on three networks and three datasets to investigate the impact of the number of trigger embedding bits. We consider the number of the chosen trigger embedding bits from 10 to 100. The experimental results are shown in Fig. \ref{fig:Trigger_Embedding_Results}. 

\begin{comment}
\begin{figure*}[]
\centering
\setcounter{subfigure}{0}
\subfloat[]{\includegraphics[width=2.2in]{figure/ResNet50-MNIST.png}}
\hspace{-1mm}
\subfloat[]{\includegraphics[width=2.2in]{figure/ResNet50-F-MNIST.png}}
\hspace{-1mm}
\subfloat[]{\includegraphics[width=2.2in]{figure/ResNet50-CIFAR.png}}
\caption{
The attack performance on ResNet50 with the different number of the trigger embedding bits ((a) MNIST, (b) F-MNIST, (c) CIFAR-10).
}
\label{fig:ResNet50}
\end{figure*}

\begin{figure*}[]
\centering
\setcounter{subfigure}{0}
\subfloat[]{\includegraphics[width=2.2in]{figure/ResNext50-MNIST.png}}
\hspace{-1mm}
\subfloat[]{\includegraphics[width=2.2in]{figure/ResNext50-F-MNIST.png}}
\hspace{-1mm}
\subfloat[]{\includegraphics[width=2.2in]{figure/ResNext50-CIFAR.png}}
\caption{
The attack performance on ResNext50 with the different number of the trigger embedding bits ((a) MNIST, (b) F-MNIST, (c) CIFAR-10).
}
\label{fig:ResNext50}
\end{figure*}

\begin{figure*}[]
\centering
\setcounter{subfigure}{0}
\subfloat[]{\includegraphics[width=2.2in]{figure/VGG16-MNIST.png}}
\hspace{-1mm}
\subfloat[]{\includegraphics[width=2.2in]{figure/VGG16-F-MNIST.png}}
\hspace{-1mm}
\subfloat[]{\includegraphics[width=2.2in]{figure/VGG16-CIFAR.png}}
\caption{
The attack performance on VGG16 with the different number of the trigger embedding bits ((a) MNIST, (b) F-MNIST, (c) CIFAR-10).
}
\label{fig:VGG16}
\end{figure*}
\end{comment}

\textbf{ResNet50}: The first row of Fig. \ref{fig:Trigger_Embedding_Results} shows the experimental results with ResNet50 on three datasets. We know that on MNIST, our \texttt{Dullahan} has little impact on the ACC and ASR keeps growing when the number of trigger embedding bits goes from 10 to 30. This phenomenon may indicate that more important bits that are strongly related to the backdoor trigger are introduced. However, when the number of the trigger embedding bits goes from 50 to 80, the ASR has a noticeable decrease. We think that more bits unrelated to the backdoor trigger may have been introduced into the trigger embedding here. When the number of trigger embedding bits goes from 80 to 100, the ASR increases obviously. We infer that the unrelated bits gradually play an unimportant role with the increase of trigger embedding bits. On F-MNIST, the number of the chosen trigger embedding bits has a certain impact on the ASR, but most of them can maintain the ASR above $80.00\%$. Moreover, compared to the baseline, there is almost no influence on the main task accuracy. On CIFAR-10, we can see that the ACC and ASR are not almost greatly affected by the number of trigger embedding bits. The ASR is kept at around $75.00\%$, and the ACC loss is within $2.20\%$.

\textbf{ResNext50}: The second row of Fig. \ref{fig:Trigger_Embedding_Results} shows the experimental results on ResNext50. For MNIST, we can see that the ACC is almost equal to the baseline, which means that our \texttt{Dullahan} has little influence on the main task. However, the ASR is not stable enough, which is reflected in the highest ASR is $99.70\%$, and the lowest ASR is nearly $60.50\%$. We suppose that it may caused by the difference between the client network and the surrogate model. Moreover, with the increase of trigger embedding bits, The bits strongly associated with the backdoor and the other unrelated bits may affect each other, resulting in an unstable ASR. For F-MNIST, the ACC is also nearly similar to the baseline. However, with the increase of trigger embedding bits, the ASR has a slight decrease and  we infer that the bits strongly related to the backdoor have been selected in the top 20 bits, and the subsequent bits, which may be less important, have a negative effect on the ASR. For CIFAR-10, both ACC and ASR maintain a steady level with the increase of trigger embedding bits. Compared with the baseline, the ACC loss is within $2.00\%$. 

\textbf{VGG16}: The last row of Fig. \ref{fig:Trigger_Embedding_Results} shows the experimental results on VGG16. We can know that both ACC and ASR have small variations with the change of the number of trigger embedding bits. Moreover, the ACC is almost equal to the baseline, which indicates that our \texttt{Dullahan} has little impact on the main task.

In summary, the experimental results show that the number of the chosen trigger embedding has some influence on the performance of our \texttt{Dullahan}. Besides, we also find that the performance of our \texttt{Dullahan} is also affected by different networks and datasets. Our method is more stable on VGG16 with the variation of the number of trigger embedding bits. We infer that the VGG16 has a simpler network structure compared with ResNet50 and ResNext50. Therefore, when executing the same epochs to produce the surrogate model, the similarity between the client network and the surrogate model on VGG16 is higher, which may lead to that the chosen trigger embedding on VGG16 is more exact. Besides, for different datasets, we find our \texttt{Dullahan} is more stable on CIFAR-10. We suppose that it may be related to our trigger patch. Note that our trigger patch is a white square with $4\times4$ pixels. In MNIST and F-MNIST datasets, the presence of numerous white pixels often results in trigger embeddings containing a higher proportion of unimportant embedding bits. We infer that the greater the difference between the backdoor feature and the normal features, the more stable the backdoor attack performance is.

\begin{figure}[htpb]
\centering
\includegraphics[width=8cm]{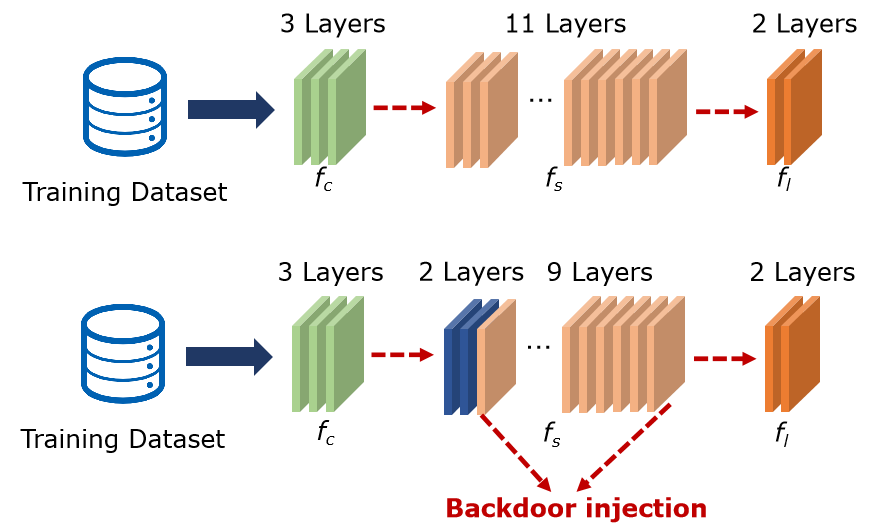}
\caption{The analysis of the splitting strategy on the attack performance.}
\label{fig:splitting strategy}
\end{figure}

\subsubsection{\textbf{Splitting Strategy}}

The existing works \cite{tajalli2023feasibility, yu2023backdoor} think that the splitting strategy is crucial in the backdoor attack for split learning because their methods need to modify the gradients to force the client network to learn the backdoor trigger. With the increase of the depth of the client network, the client network can learn less information about the backdoor trigger.

However, different from the existing backdoor attack methods, our method conducts a backdoor attack by injecting trigger embedding into the server network, which just needs to collect the relevant training data during the training process. That is, our method does not require modifying the client network by manipulating the training process. Moreover, because the attacker knows the architecture of the client network, he/she can choose which layer to inject trigger embedding. Therefore, the splitting strategy is not sensitive to the attack performance of our method. We give an example to show the reasons in detail. Specifically, we assume that there is a split learning with VGG16 and the client network is $3$ layers, the server network is $11$ layers, and the last network is $2$ layers, as shown in Fig. \ref{fig:splitting strategy}. In our \texttt{Dullahan}, the attacker can obtain all the inputs and the trained server network during the training process. The corresponding outputs can be also calculated based on the collected inputs and server network. The attacker can also inject trigger embedding into the last $9$ layers of the server network based on the inputs and outputs of these $9$ layers rather than modify the whole server network. Though there are $3$ layers in the client network, the attacker can also inject a backdoor in the fifth layer of the model rather than the third layer. Hence, in our \texttt{Dullahan}, the attacker just needs to inject the trigger embedding in the layer, which is more than the number of the client network layers. The attack performance is little influenced by the splitting strategy.

\begin{figure*}[htp]
\centering
\includegraphics[width=18cm]{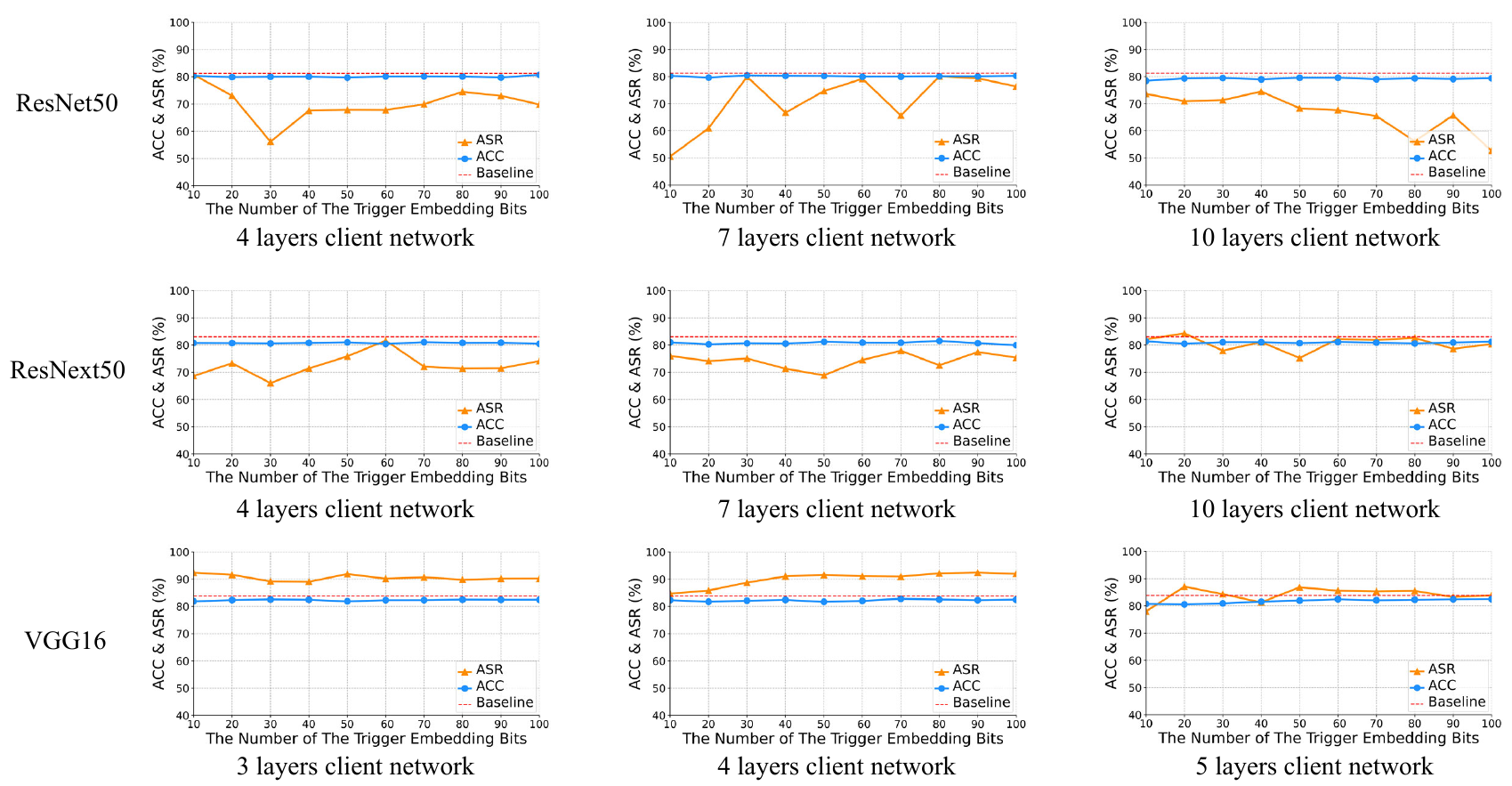}
\caption{The attack performance on three models and three datasets under various splitting strategies. The figures from top to bottom row are the experimental results on ResNet50, ResNext50, and VGG16, respectively. The figures from left to right column are the experimental results of different splitting strategies. Note that three splitting strategies, including the client network with 4, 7, and 10 layers, are considered in ResNet50 and ResNext50. Three splitting strategies, including the client network with 3, 4, and 5 layers, are considered in VGG16.}
\label{fig:splitting_strategy_results}
\end{figure*}

\begin{comment}
\begin{figure*}[htbp]
\centering
\setcounter{subfigure}{0}
\subfloat[]{\includegraphics[width=2.2in]{figure/ResNet50_4.png}}
\hspace{-1mm}
\subfloat[]{\includegraphics[width=2.2in]{figure/ResNet50_7.png}}
\hspace{-1mm}
\subfloat[]{\includegraphics[width=2.2in]{figure/ResNet50_10.png}}
\caption{
The attack performance on ResNet50 with the various splitting strategies ((a)4 layers client network, (b)7 layers client network, (c)10 layers client network).
}
\label{fig:ResNet50-2}
\end{figure*}

\begin{figure*}[htbp]
\centering
\setcounter{subfigure}{0}
\subfloat[]{\includegraphics[width=2.2in]{figure/ResNext50_4.png}}
\hspace{-1mm}
\subfloat[]{\includegraphics[width=2.2in]{figure/ResNext50_7.png}}
\hspace{-1mm}
\subfloat[]{\includegraphics[width=2.2in]{figure/ResNext50_10.png}}
\caption{
The attack performance on ResNext50 with the various splitting strategies ((a)4 layers client network, (b)7 layers client network, (c)10 layers client network).
}
\label{fig:ResNext50-2}
\end{figure*}

\begin{figure*}[htbp]
\centering
\setcounter{subfigure}{0}
\subfloat[]{\includegraphics[width=2.2in]{figure/VGG16_3.png}}
\hspace{-1mm}
\subfloat[]{\includegraphics[width=2.2in]{figure/VGG16_4.png}}
\hspace{-1mm}
\subfloat[]{\includegraphics[width=2.2in]{figure/VGG16_5.png}}
\caption{
The attack performance on VGG16 with the various splitting strategies ((a)3 layers client network, (b)4 layers client network, (c)5 layers client network).
}
\label{fig:VGG16-2}
\end{figure*}
\end{comment}

\subsection{Attacker with No Knowledge of The Client Network}
Sometimes, it may be difficult for the attacker to infer the detailed architecture of the client network in a real-world scenario. Therefore, in this subsection, we experimentally demonstrate that our \texttt{Dullahan} can also achieve a fine attack performance when the attacker has no knowledge about the client network. 

Due to the lack of knowledge about the client network, the attacker can build a different architecture surrogate model whose architecture is shown in Appendix Table \ref{tab:architecture}. We demonstrate the effectiveness of the backdoor attack on CIFAR-10 and three models whose splitting strategies are shown in Appendix Table \ref{tab:st}. It is worth mentioning that due to the varying sizes of different layers' interfaces, the selection of surrogate models for different neural networks and splitting strategies may also differ slightly. For the Split 2 and Split 3 splitting strategies in the VGG16, we employ the Surrogate model-2 and Surrogate model-3 to approximate the client network. For the other neural networks and splitting strategies, we use the Surrogate model-1 to approximate the client network. 

We also analyze the influence of the number of trigger embedding bits on the attack performance. In the experiment, we inject backdoor trigger embedding into the server network for a total of 20 epochs and record the mean ACC and ASR for executing 16 epochs to 20 epochs as the final results. Moreover, all the other experimental settings are the same as Subsection \ref{label:setting}. The experimental results are shown in Fig. \ref{fig:splitting_strategy_results}. We know that our method is still effective when
having no knowledge about the client network. Then, we will give a detailed analysis of the experimental results of each network as follows.

\textbf{ResNet50}: Owing to the residual structure in ResNet50, we consider three splitting strategies with the client network having 4, 7, and 10 layers, respectively. Compared with knowing the architecture of the client network (The third column in the first row of Fig. \ref{fig:Trigger_Embedding_Results}), both the ACC and ASR have a slight decrease. Specifically, the ACC loss is within $3.00\%$ compared to the baseline and the ASR is around $70.00\%$. When the architecture of the client network is unknown, we can infer that the surrogate model will differ more from the client network. Moreover, from the first row of Fig. \ref{fig:splitting_strategy_results}, we can see that the proposed backdoor method is not sensitive to the splitting strategies. Besides, similar to the third column in the first row of Fig. \ref{fig:Trigger_Embedding_Results}, with the difference in the number of trigger embedding bits, the ASR also has a certain fluctuation due to the differentiation between the client network and the surrogate model.

\textbf{ResNext50}: Similar to ResNet50, we also take into account three splitting strategies that the client network has 4, 7, and 10 layers, respectively. Compared with knowing the architecture of the client network (The third column in the second row of Fig. \ref{fig:Trigger_Embedding_Results}), both the ACC and ASR have almost no decrease. Compared with the baseline, the ACC loss is within $2.60\%$ and the ASR is around $75.00\%$. Moreover, we infer that the ASR is also slightly influenced by the number of trigger embedding bits, due to the slight increase of the ASR with the increase of the client network layers.

\textbf{VGG16}: For VGG16, we also consider three splitting strategies that the client network has 3, 4, and 5 layers, respectively. Compared with knowing the architecture of the client network (The third column in the third row of Fig. \ref{fig:Trigger_Embedding_Results}), the ASR has a small increase, and the ACC loss (within $3.30\%$) is almost the same as the third column in the first row of Fig. \ref{fig:Trigger_Embedding_Results}. Moreover, with the increase of the client network layers, both the ASR and ACC have little change. Besides, the number of trigger embedding bits also has little influence on the ACC and ASR.

\begin{table*}[htp]
    \caption{Defense against the proposed \texttt{Dullahan}.}
    \label{tab:defense result}
    \centering
    \begin{tabular}{ccccccc}
        \midrule
         \thead{Noise\\Scale} & Model & Baseline & \thead{ACC\\(Adding Noise)} & \thead{ACC\\(Adding Noise \\ and Backdoor)} & \thead{Original\\ASR} & \thead{ASR\\(Adding Noise)}\\
        \midrule
        \multirow{3}{*}{0.05} & ResNet50 & 81.30 & 77.06 & 76.13 & 75.40 & 75.40\\
         & ResNext50 & 83.00 & 80.80 & 79.31 & 79.40 & 70.60  \\
         & VGG16 & 83.80 & 82.16 & 80.73 & 81.40 & 86.50 \\
        \midrule
        \multirow{3}{*}{0.10} & ResNet50 & 81.30 & 74.30 & 72.80 &  75.40 & 83.30  \\
         & ResNext50 & 83.00 & 77.01 & 75.32 & 79.40 & 68.50   \\
         & VGG16 & 83.80 & 77.37 & 76.91 & 81.40 & 77.80 \\
        \midrule
    \end{tabular}
\end{table*}

\subsection{Defense Evaluation}

In this section, to further evaluate the performance of the proposed \texttt{Dullahan}, we assume that the client side may deploy some potential defense methods, such as adding Gaussian noise to the training data. Therefore, we plan to evaluate the performance of the presented \texttt{Dullahan} by employing the Gaussian noise defense. We conduct experiments on CIFAR-10 and three models with the same settings as in Subsection \ref{label:setting}. We evaluate the defense performance against the proposed \texttt{Dullahan} on two scales of Gaussian noise. The experimental results are reported in Table \ref{tab:defense result}. It can be seen that adding Gaussian noise cannot mitigate the risk of the proposed \texttt{Dullahan}.

Specifically, with the increase of the Gaussian noise scale, the main task ACC drops significantly. For ResNet50, when the noise scales are $0.05$ and $0.10$, compared with the baseline $81.30\%$, the ACC is $77.06\%$ and $74.30\%$, respectively. Secondly, after employing the defense method in the training process, our \texttt{Dullahan} has a slight impact on the main task. For example, for VGG16, when the noise scale is $0.10$, the ACC just drops $0.46\%$ after injecting the backdoor. Thirdly, with the defense method, the performance of our \texttt{Dullahan} fluctuates slightly. However, It generally maintains a satisfactory attack performance. For instance, when the noise scale is $0.05$, for ResNet50, the ASR remains stable. For ResNext50, the ASR drops $8.80\%$. For VGG16, the ASR increases $5.10\%$. Finally, we can find that the noise scale has little influence on the performance of our \texttt{Dullahan}. For example, for ResNext50, when the noise scale is $0.05$ and $0.10$, the ACC is $79.31\%$ and $75.32\%$, respectively. The ASR is $70.60\%$ and $68.50\%$, respectively. In conclusion, adding Gaussian noise to the training data has a significant influence on the main task and cannot mitigate the risk of the proposed \texttt{Dullahan}.

\section{Conclusion}
In this paper, we first propose a Stealthy Backdoor Attack Strategy (namely \texttt{Dullahan}) tailored to the without-label-sharing split learning architecture. In the presented attack method, the attacker can inject a backdoor trigger just by modifying the server network after finishing the standard training process. Moreover, the extensive experiments also validate that the proposed \texttt{Dullahan} method can achieve the satisfactory ASR no matter whether the attacker knows the architecture of the client network. The ACC is also nearly close to the baseline. Because there is no modification on the intermediate parameters (e.g., gradients) in our \texttt{Dullahan}, it is difficult for the client side to detect backdoor injection during the training process by monitoring the intermediate parameters. Therefore, for the client side, when employing the split learning architecture, it is necessary to detect the security vulnerability of the server network after finishing training. Besides, to guarantee the security of split learning, it is also essential to design a defense strategy that can maintain the main task performance and destroy the attack performance. In conclusion, our method exposes the vulnerability of splitting learning, which can also promote the development of relevant defense technologies. 

%Bibliography
\bibliographystyle{IEEEtran}
\bibliography{main}  

\section{Appendix}
For knowing the client network's architecture, Table \ref{tab:networks} shows the splitting strategies of three models. For having no knowledge about the client network's architecture, Table \ref{tab:architecture} shows the architecture of the surrogate model. Table \ref{tab:st} shows the splitting strategies of $3$ models.

\begin{table}[htp]
    \caption{The architecture of the surrogate model.}
    \label{tab:architecture}
    \centering
    \begin{tabular}{ccccc}
        \midrule
             & Architecture & \\
        \midrule
        \thead{
        Surrogate model-1\\
        (256 * 8 * 8)
        } & \thead{
        Conv2d(in\_channels, 64, (3 3), 2, 1)\\
        BN (64), ReLU()\\
        Conv2d(64, 128, (3 3), 2, 1)\\
        BN (128), ReLU()\\
        Conv2d(128, 256, (3 3), 1, 1)\\
        BN (256), ReLU()\\
        Residual\_block(256)\\
        Residual\_block(256)\\
        Residual\_block(256)\\
        Residual\_block(256)\\
        Residual\_block(256)\\
        Residual\_block(256)\\
        } & \\
        \midrule
        \thead{
        Surrogate model-2\\
        (128 * 16 * 16)
        } & \thead{
        Conv2d(in\_channels, 64, (3 3), 2, 1)\\
        BN (64), ReLU()\\
        Conv2d(64, 128, (3 3), 1, 1)\\
        BN (128), ReLU()\\
        Residual\_block(128)\\
        Residual\_block(128)\\
        Residual\_block(128)\\
        Residual\_block(128)\\
        Residual\_block(128)\\
        Residual\_block(128)\\
        } & \\
        \midrule
        \thead{
        Surrogate model-3\\
        (128 * 8 * 8)
        } & \thead{
        Conv2d(in\_channels, 64, (3 3), 2, 1)\\
        BN (64), ReLU()\\
        Conv2d(64, 128, (3 3), 2, 1)\\
        BN (128), ReLU()\\
        Residual\_block(128)\\
        Residual\_block(128)\\
        Residual\_block(128)\\
        Residual\_block(128)\\
        Residual\_block(128)\\
        Residual\_block(128)\\
        } & \\
        \midrule
    \end{tabular}
\end{table}

\begin{table*}[htp]
    \caption{The networks and split strategies with knowing the client network's architecture.}
    \label{tab:networks}
    \centering
    \begin{tabular}{ccccc}
        \midrule
        \textbf{ Model }& \textbf{$f_c$} & \textbf{$f_s$} & $f_l$\\
        \midrule
        \multirow{1}{*}{ResNet50} & \thead{
        Conv2d(in\_channels, 64, (7, 7), 2, 3)\\
        BN (64), ReLU()\\
        MaxPool((3,3), 2, 1)\\
        Stage1
        } & \thead{
        Stage2\\
        Stage3\\
        Stage4\\
        MaxPool((1,1))\\
        } & \thead{Linear(2048, num\_classes)}\\
        \midrule
        \multirow{1}{*}{ResNext50} & \thead{
        Conv2d(in\_channels, 64, (7, 7), 2, 3)\\
        BN (64), ReLU()\\
        MaxPool((3,3), 2, 1)\\
        Stage1
        } & \thead{
        Stage2\\
        Stage3\\
        Stage4\\
        MaxPool((1,1))\\
        }  & \thead{Linear(2048, num\_classes)}\\
        \midrule
        \multirow{1}{*}{VGG16} & \thead{
        Conv2d(in\_channels, 64, (3 3), 1, 1)\\
        BN (64), ReLU()\\
        Conv2d(64, 64, (3 3), 1, 1)\\
        BN(64), ReLU()\\
        MaxPool((2,2), 2)\\
        Conv2d(64, 128, (3 3), 1, 1)\\
        BN (128), ReLU()\\
        Conv2d(128, 128, (3 3), 1, 1)\\
        BN(128), ReLU()\\
        MaxPool((2,2), 2)\\
        Conv2d(128, 256, (3 3), 1, 1)\\ 
        BN (256), ReLU()\\
        } & \thead{
        Conv2d(256, 256, (3 3), 1, 1)\\
        BN(256), ReLU()\\
        ...\\
        MaxPool((2,2), 2)\\
        MaxPool((7,7))\\
        }  & \thead{Linear(25088, num\_classes)}\\
        \midrule
    \end{tabular}
\end{table*}

\begin{table*}[h]
    \caption{The networks and split strategies without knowing the client network's architecture.}
    \label{tab:st}
    \centering
    \begin{tabular}{ccccc}
        \midrule
        \textbf{ Model }& Split Strategies & \textbf{$f_c$} & \textbf{$f_s$} & $f_l$\\
        \midrule
        \multirow{11}{*}{ResNet50} & \thead{Split 1} & \thead{
        Conv2d(in\_channels, 64, (7, 7), 2, 3)\\
        BN (64), ReLU()\\
        MaxPool((3,3), 2, 1)\\
        Conv\_block(64, 256, groups = 1)
        } & \thead{
        Conv\_block(256, 256, groups = 1)\\
        Conv\_block(256, 256, groups = 1)\\
        Stage2\\
        Stage3\\
        Stage4\\
        MaxPool((1,1))\\
        } & \thead{Linear(2048, num\_classes)}\\
        & \thead{Split 2} & \thead{
        Conv2d(in\_channels, 64, (7, 7), 2, 3)\\
        BN (64), ReLU()\\
        MaxPool((3,3), 2, 1)\\
        Conv\_block(64, 256, groups = 1)\\
        Conv\_block(256, 256, groups = 1)
        } & \thead{
        Conv\_block(256, 256, groups = 1)\\
        Stage2\\
        Stage3\\
        Stage4\\
        MaxPool((1,1))\\
        } & \thead{Linear(2048, num\_classes)}\\
        & \thead{Split 3} & \thead{
        Conv2d(in\_channels, 64, (7, 7), 2, 3)\\
        BN (64), ReLU()\\
        MaxPool((3,3), 2, 1)\\
        Stage1
        } & \thead{
        Stage2\\
        Stage3\\
        Stage4\\
        MaxPool((1,1))\\
        } & \thead{Linear(2048, num\_classes)}\\
        \midrule
        \multirow{11}{*}{ResNext50} & Split 1 & \thead{
        Conv2d(in\_channels, 64, (7, 7), 2, 3)\\
        BN (64), ReLU()\\
        MaxPool((3,3), 2, 1)\\
        Conv\_block(64, 256, groups = 32)
        } & \thead{
        Conv\_block(256, 256, groups = 32)\\
        Conv\_block(256, 256, groups = 32)\\
        Stage2\\
        Stage3\\
        Stage4\\
        MaxPool((1,1))\\
        }  & \thead{Linear(2048, num\_classes)}\\
        & Split 2 & \thead{
        Conv2d(in\_channels, 64, (7, 7), 2, 3)\\
        BN (64), ReLU()\\
        MaxPool((3,3), 2, 1)\\
        Conv\_block(64, 256, groups = 32)\\
        Conv\_block(256, 256, groups = 32)
        } & \thead{
        Conv\_block(256, 256, groups = 32)\\
        Stage2\\
        Stage3\\
        Stage4\\
        MaxPool((1,1))\\
        }  & \thead{Linear(2048, num\_classes)}\\
        & Split 3 & \thead{
        Conv2d(in\_channels, 64, (7, 7), 2, 3)\\
        BN (64), ReLU()\\
        MaxPool((3,3), 2, 1)\\
        Stage1
        } & \thead{
        Stage2\\
        Stage3\\
        Stage4\\
        MaxPool((1,1))\\
        }  & \thead{Linear(2048, num\_classes)}\\
        \midrule
         \multirow{23}{*}{VGG16} & Split 1 & \thead{
        Conv2d(in\_channels, 64, (3 3))\\
        BN (64), ReLU()\\
        Conv2d(64, 64, (3 3))\\
        BN(64), ReLU()\\
        MaxPool((2,2), 2)\\
        Conv2d(64, 128, (3 3))\\
        BN (128), ReLU()\\
        } & \thead{
        Conv2d(128, 128, (3 3), 1, 1)\\
        BN(128), ReLU()\\
        MaxPool((2,2), 2)\\
        Conv2d(128, 256, (3 3), 1, 1)\\
        BN(256), ReLU()\\
        ...\\
        MaxPool((2,2), 2)\\
        MaxPool((7,7))\\
        }  & \thead{Linear(25088, num\_classes)}\\
        & Split 2 & \thead{
        Conv2d(in\_channels, 64, (3 3), 1, 1)\\
        BN (64), ReLU()\\
        Conv2d(64, 64, (3 3), 1, 1)\\
        BN(64), ReLU()\\
        MaxPool((2,2), 2)\\
        Conv2d(64, 128, (3 3), 1, 1)\\
        BN (128), ReLU()\\
        Conv2d(128, 128, (3 3), 1, 1)\\
        BN(128), ReLU()\\
        MaxPool((2,2), 2)\\
        } & \thead{
        Conv2d(128, 256, (3 3), 1, 1)\\
        BN(256), ReLU()\\
        Conv2d(256, 256, (3 3), 1, 1)\\
        BN(256), ReLU()\\
        ...\\
        MaxPool((2,2), 2)\\
        MaxPool((7,7))\\     
        }  & \thead{Linear(25088, num\_classes)}\\
        & Split 3 & \thead{
        Conv2d(in\_channels, 64, (3 3), 1, 1)\\
        BN (64), ReLU()\\
        Conv2d(64, 64, (3 3), 1, 1)\\
        BN(64), ReLU()\\
        MaxPool((2,2), 2)\\
        Conv2d(64, 128, (3 3), 1, 1)\\
        BN (128), ReLU()\\
        Conv2d(128, 128, (3 3), 1, 1)\\
        BN(128), ReLU()\\
        MaxPool((2,2), 2)\\
        Conv2d(128, 256, (3 3), 1, 1)\\ 
        BN (256), ReLU()\\
        } & \thead{
        Conv2d(256, 256, (3 3), 1, 1)\\
        BN(256), ReLU()\\
        ...\\
        MaxPool((2,2), 2)\\
        MaxPool((7,7))\\
        }  & \thead{Linear(25088, num\_classes)}\\
        \midrule
    \end{tabular}
\end{table*}

\end{document}